\documentclass[letterpaper,12pt]{article}
\usepackage{amsmath}
\usepackage{amstext}
\usepackage{graphicx}
\usepackage{amsfonts}
\usepackage{amssymb}
\begin{document}

\author{Michael Frankel\\{\small \textit{Department of Mathematical Sciences,}}\\{\small \textit{Indiana University Purdue University Indianapolis, }}\\{\small \textit{\ Indianapolis,}} {\small \textit{\ IN 46202-3216 U.S.A. }}\\{\small \textit{email: mfrankel@math.iupui.edu}}
\and Victor Roytburd\thanks{Corresponding author; Phone 518-276-6889, Fax 518-276-4824}\\{\small \textit{Department of Mathematical Sciences, }}\\{\small \textit{\ Rensselaer Polytechnic Institute, }}\\{\small \textit{\ Troy, NY 1218-3590, U.S.A. }}\\{\small \textit{email:roytbv@rpi.edu}}}
\title{Frequency Locking for Combustion Synthesis in Periodic
Medium{\normalsize \textbf{\ }}}
\date{\empty}
\maketitle
\begin{abstract}
Solutions of a 1-D free-interface problem modeling solid combustion front
propagating in combustible mixture with periodically varying concentration of
reactant exhibit classical phenomenon of mode locking. Numerical simulation
shows a variety of locked periodic, \ quasi-periodic and chaotic solutions.

\textbf{PACS}: 05.45.-a, 02.30.Oz, 81.20.Ka

\textbf{Keywords}: Free-interface problems, periodic medium, mode locking,
Arnold's tongues
\end{abstract}

%Started December, 2003
%
%

\section{Introduction}

The current letter is intended to communicate new observations based on
numerical solutions of the two-phase Stefan problem with kinetics. The
simulations reveal some previously unknown features of this dynamically
diverse system. Namely, it exhibits the phenomenon of frequency locking in
response to a spatially-periodic perturbation of the medium.

The free-boundary problem that is the subject of the paper\ arises naturally
as a mathematical model of a variety of exothermic phase transition type
processes, such as condensed-state combustion (also known as Self-propagating
High-temperature Synthesis or SHS \cite{munir,var2}),{\normalsize \ }
solidification with undercooling \cite{langer}, laser induced evaporation
\cite{gt}, rapid crystallization in thin films \cite{VanWee} etc.

This study represents a natural extension of the numerical experiments
described in \cite{ctm-port2}, where it was demonstrated that, due to the
competition between the heat release at the interface and the heat dissipation
by the medium, the system generates a variety of complex thermokinetic
oscillations. The dynamical patterns exhibited by the unperturbed system, as
the governing parameters are varied, include Hopf bifurcation, period doubling
cascades leading to chaotic pulsations, Shilnikov-Hopf bifurcation etc.

The new feature that was added to the setting is a variable initial
concentration of the so-called deficient component which controls the reaction
rate in the case of the combustion synthesis. We assume the concentration to
be a periodically perturbed constant. While the original objective of our
experiments was to study the dependence of the mean propagation velocity on
the perturbation frequency and amplitude, we have noticed that the dynamics
was qualitatively substantially different from that of the unperturbed
problem. What we observed was a variety of quasi-periodic and complex periodic
regimes with periods that are various multiples of the perturbation period.
These observations present a convincing evidence that we are dealing here with
the phenomenon of frequency locking.

The physical phenomena modeled by periodically driven dynamical systems appear
in many fields (see e.g. \cite{verhulst}) such as lasers, superconductors
(Josephson's junctions), mechanical engineering, etc. Theoretical as well as
numerical studies show that periodic forcing can drive these systems to
exhibit rich patterns of behavior that includes mode loking, with the
mode-locked bands usually having the structure of the so-called Arnold's
tongues in the amplitude-frequency parameter space.

It is necessary to mention that the analysis or numerical simulation or a
combination of thereof is usually carried out for a finite-dimensional system
even if the original physical system is infinite-dimensional. The reduction in
such cases is based on an assumption of a certain ansatz which consequently
leads to an ODE. At the same time the perturbation is usually external to the
base system and it is periodic in time. In our case, however, we make no
attempt of such reduction, neither are we aware of any ansatz for a solution
leading to it, while \textit{the internal forcing is} \textit{periodic in}
\textit{space}, and the system remains formally autonomous.

Below we present some examples of numerical solutions and a crude map of a
basic resonance band in the amplitude-frequency parameter space that
represents an Arnold's tongue. We remind the reader that every point in the
parameter space is a result of numerical simulation of a nontrivial
free-interface problem for a partial differential equation.

Thus, the sharp interface model of combustion synthesis presents a very
natural and transparent example of the frequency locking phenomenon for a PDE.

\section{Free-interface problem{\protect\normalsize \textbf{\ }}}

In the context of combustion of condensed matter a wave of exothermic chemical
reaction transforms a solid combustible mixture directly into solid product.
The 1-D model involves differential equations for the temperature $u $ of the
mixture and the relative concentration of the so-called deficient reactant $Z$
(see e.g. Shkadinsky \textit{et al. }\cite{ShkaKhaMer}):
\begin{align*}
u_{t}  &  =\kappa u_{xx}+qW(Z,u),\\
Z_{t}  &  =-W(Z,u),
\end{align*}
where $\kappa$ is the thermal diffusivity, $W$ is the chemical reaction rate,
and $q$ is the heat release.

Due to a strong temperature dependence of the reaction there is a well defined
narrow region (flame front) where the bulk of chemical reaction and the heat
release occur. Thus the distributed chemical reaction can be replaced by the
$\delta$-function (see Zeldovich \textit{et al.} \cite{ZelBarLibMak}),
\[
W=Z(s(t))g(u)\delta(x-s(t))
\]
located at the interface $x=s(t)$ between the fresh, $Z=Z(s(t))$ and burned
$Z=0$ material.

The equation with the $\delta$-function source is rewritten as a system of two
heat equations coupled at the interface. In the context of solidification with
undercooling \cite{langer} or rapid crystallization of thin films
\cite{VanWee}, the free-interface model studied below is conceptually even
simpler: the latent heat of the phase transition released at the interface
must be diffused by the surrounding matter.

Therefore we shall be concerned with the following \textit{appropriately
non-dimensionalized} free-interface problem: find $s(t)$ and $u(x,t)$ such
that
\begin{gather}
u_{t}=u_{xx},\quad x\neq s(t),\label{he}\\
u(x,0)=u_{0}(x)\geq0,\label{ic}\\
Z_{0}(s(t))g[u(s(t),t)]=v(t)\quad\mathrm{for}\,t>0,\label{kc}\\
u_{x}^{+}(s(t),t)-u_{x}^{-}(s(t),t)=v(t)\quad\mathrm{for}\,t>0, \label{jc}%
\end{gather}
where $v(t)$ is the interface velocity, $s(t)=\int_{0}^{t}v(\tau)d\tau$ is its
position, $u$ is the temperature, and the derivatives $u_{x}^{+}$ and
$u_{x}^{-}$ are taken from right side and left side of the free interface
respectively. At $-\infty$ the surrounding matter is assumed to be at the
temperature of the fresh combustible, while the temperature of the burned
matter as well as its gradient is bounded:
\begin{equation}
u(-\infty,t)=0,\;u(\infty,t)<C,\;|u_{x}(\infty,t)|<C. \label{b.c.}%
\end{equation}

Under reasonable assumptions on the kinetics functions $g,$ one can rigorously
prove that the free-interface problem (\ref{he})-(\ref{jc}) possesses global
in time, uniformly bounded classical solutions (see \cite{poly-long}). One
should not regard the rigorous proof of the existence and uniform boundedness
of solutions as a futile academic exercise but rather view it as a
verification of correctness of the model. It provides a firm foundation for
the numerical simulations presented in the paper. In particular, the proof of
uniform boundedness of solutions underscores the dynamical robustness of the model.

\section{Overview of dynamics for unforced problem}

For reader's convenience in this Section we give a brief overview of the
unforced dynamics (see \cite{ctm-port2}), i.e., with $Z_{0}(x)\equiv1$. It is
convenient to rewrite the non-equilibrium interface condition (\ref{kc}) in
the form:
\begin{equation}
v=g[u(s(t),t)]:=1+\alpha J(u(s(t),t)) \label{disp8}%
\end{equation}
We shall assume that the function $J(\xi)=(g(\xi)-1)/\alpha$ is normalized in
such a way that
\begin{equation}
J(1)=0,~~~~~J^{\prime}(1)=-1, \label{scale}%
\end{equation}
which can be achieved by rescaling variables. This assumption makes the
problems with different kinetics identical in terms of linearization about the
basic solution. We note that the variables are selected so that the
\textit{interface propagates to the left}.

In order to clarify the meaning of the (positive) parameter $\alpha$ which is
the main instability parameter, we note that for the Arrhenius type kinetics
\begin{equation}
v=g(u)=-\exp[\frac{\alpha(u-1)}{\sigma+(1-\sigma)u}] \label{arh1}%
\end{equation}
$\alpha$ is the scaled activation energy for the exothermic chemical reaction
that occurs at the interface, and $\sigma$ is the temperature ratio of the
fresh and burned material for the traveling wave solution.

The problem (\ref{he})-(\ref{jc}) has a unique traveling wave solution
\begin{equation}
u_{b}=\{%
%TCIMACRO{\QATOP{exp(x+t),\quad x\leq-t}{\quad1,\quad\quad\quad x>-t}}%
%BeginExpansion
\genfrac{}{}{0pt}{}{exp(x+t),\quad x\leq-t}{\quad1,\quad\quad\quad x>-t}%
%EndExpansion
,~~~~s_{b}=-t. \label{disp9}%
\end{equation}
provided $J$ is monotone. The linear stability analysis indicates that the
loss of stability occurs via a supercritical Hopf bifurcation at $\alpha
_{cr}=\sqrt{5}+2$, the corresponding frequency is $\omega_{cr}\simeq1.03,$
$T_{cr}=2\pi/\omega_{cr}\simeq6.1.$%
%TCIMACRO{\TeXButton{B}{\begin{table}[tbp]\centering}}%
%BeginExpansion
\begin{table}[tbp]\centering
%EndExpansion%
\begin{tabular}
[c]{|llllllllllllllllllll|}\hline
& $\bigtriangledown$ & $\bigtriangledown$ &  & $\bigtriangledown$ &
$\bigtriangledown$ & $\bigtriangledown$ & $\bigtriangledown$ &
$\bigtriangledown$ & $\bigtriangledown$ & $\bigtriangledown$ &
$\bigtriangledown$ & $\bigtriangledown$ & $\bigtriangledown$ &
$\bigtriangledown$ & $\bigtriangledown$ & $\bigtriangledown$ &
$\bigtriangledown$ & $\bigtriangledown$ & $\bigtriangledown$\\
&  & $\bigtriangledown$ &  & $\bigtriangledown$ & $\bigtriangledown$ &
$\bigtriangledown$ & $\bigtriangledown$ & $\bigtriangledown$ &
$\bigtriangledown$ & $\bigtriangledown$ & $\bigtriangledown$ &
$\bigtriangledown$ & $\bigtriangledown$ & $\bigtriangledown$ &
$\bigtriangledown$ & $\bigtriangledown$ & $\bigtriangledown$ &
$\bigtriangledown$ & $\bigtriangledown$\\
&  & $\bigtriangledown$ &  & $\bigtriangledown$ & $\bigtriangledown$ &
$\bigtriangledown$ & $\bigtriangledown$ & $\bigtriangledown$ &
$\bigtriangledown$ & $\bigtriangledown$ & $\bigtriangledown$ &
$\bigtriangledown$ & $\bigtriangledown$ & $\bigtriangledown$ &
$\bigtriangledown$ & $\bigtriangledown$ & $\bigtriangledown$ &
$\bigtriangledown$ & $\bigtriangledown$\\
&  & $\bigtriangledown$ &  & $\bigtriangledown$ &  & $\bigtriangledown$ &
$\bigtriangledown$ & $\bigtriangledown$ & $\bigtriangledown$ &
$\bigtriangledown$ & $\bigtriangledown$ & $\bigtriangledown$ &
$\bigtriangledown$ & $\bigtriangledown$ & $\bigtriangledown$ &
$\bigtriangledown$ & $\bigtriangledown$ & $\bigtriangledown$ &
$\bigtriangledown$\\
&  & $\bigtriangledown$ &  & $\bigtriangledown$ &  & $\bigtriangledown$ &
$\bigtriangledown$ & $\bigtriangledown$ & $\bigtriangledown$ &
$\bigtriangledown$ & $\bigtriangledown$ & $\bigtriangledown$ &
$\bigtriangledown$ & $\bigtriangledown$ & $\bigtriangledown$ &
$\bigtriangledown$ & $\bigtriangledown$ & $\bigtriangledown$ &
$\bigtriangledown$\\
&  & $\bigtriangledown$ &  & $\bigtriangledown$ &  & $\bigtriangledown$ &
$\bigtriangledown$ & $\bigtriangledown$ & $\bigtriangledown$ &
$\bigtriangledown$ & $\bigtriangledown$ & $\bigtriangledown$ &
$\bigtriangledown$ & $\bigtriangledown$ & $\bigtriangledown$ &
$\bigtriangledown$ & $\bigtriangledown$ & $\bigtriangledown$ &
$\bigtriangledown$\\
&  & $\bigtriangledown$ &  & $\bigtriangledown$ &  & $\bigtriangledown$ &
$\bigtriangledown$ & $\bigtriangledown$ & $\bigtriangledown$ &
$\bigtriangledown$ & $\bigtriangledown$ & $\bigtriangledown$ &
$\bigtriangledown$ & $\bigtriangledown$ & $\bigtriangledown$ &
$\bigtriangledown$ & $\bigtriangledown$ & $\bigtriangledown$ &
$\bigtriangledown$\\
&  & $\bigtriangledown$ &  & $\bigtriangledown$ &  & $\bigtriangledown$ &
$\bigtriangledown$ & $\bigtriangledown$ & $\bigtriangledown$ &
$\bigtriangledown$ & $\bigtriangledown$ & $\bigtriangledown$ &
$\bigtriangledown$ & $\bigtriangledown$ & $\bigtriangledown$ &
$\bigtriangledown$ & $\bigtriangledown$ & $\bigtriangledown$ & \\
&  &  &  &  &  & $\bigtriangledown$ & $\bigtriangledown$ & $\bigtriangledown$%
& $\bigtriangledown$ & $\bigtriangledown$ & $\bigtriangledown$ &
$\bigtriangledown$ & $\bigtriangledown$ & $\bigtriangledown$ &
$\bigtriangledown$ & $\bigtriangledown$ & $\bigtriangledown$ &
$\bigtriangledown$ & \\
&  &  &  &  &  &  & $\bigtriangledown$ & $\bigtriangledown$ &
$\bigtriangledown$ & $\bigtriangledown$ & $\bigtriangledown$ &
$\bigtriangledown$ & $\bigtriangledown$ & $\bigtriangledown$ &
$\bigtriangledown$ & $\bigtriangledown$ & $\bigtriangledown$ &
$\bigtriangledown$ & \\
&  &  &  &  &  &  & $\bigtriangledown$ & $\bigtriangledown$ &
$\bigtriangledown$ & $\bigtriangledown$ & $\bigtriangledown$ &
$\bigtriangledown$ & $\bigtriangledown$ & $\bigtriangledown$ &
$\bigtriangledown$ & $\bigtriangledown$ &  & $\bigtriangledown$ & \\
&  &  &  &  &  &  &  & $\bigtriangledown$ & $\bigtriangledown$ &
$\bigtriangledown$ & $\bigtriangledown$ & $\bigtriangledown$ &
$\bigtriangledown$ & $\bigtriangledown$ & $\bigtriangledown$ &  &  &  &
$\bigtriangledown$\\
&  &  &  &  &  &  &  & $\bigtriangledown$ & $\bigtriangledown$ &
$\bigtriangledown$ & $\bigtriangledown$ & $\bigtriangledown$ &
$\bigtriangledown$ & $\bigtriangledown$ &  &  &  &  & $\bigtriangledown$\\
&  &  &  &  &  &  &  &  & $\bigtriangledown$ & $\bigtriangledown$ &
$\bigtriangledown$ & $\bigtriangledown$ & $\bigtriangledown$ &
$\bigtriangledown$ &  &  &  &  & $\bigtriangledown$\\
&  &  &  &  &  &  &  &  & $\bigtriangledown$ & $\bigtriangledown$ &
$\bigtriangledown$ & $\bigtriangledown$ & $\bigtriangledown$ &  &  &  &  &  &
\\
&  &  &  &  &  &  &  &  & $\bigtriangledown$ & $\bigtriangledown$ &
$\bigtriangledown$ & $\bigtriangledown$ & $\bigtriangledown$ &  &  &  &  &  &
\\
&  &  &  &  &  &  &  &  &  & $\bigtriangledown$ & $\bigtriangledown$ &
$\bigtriangledown$ &  &  &  &  &  &  & \\
&  &  &  &  &  &  &  &  &  & $\bigtriangledown$ & $\bigtriangledown$ &  &  &
&  &  &  &  & \\
&  &  &  &  &  &  &  &  &  & $\bigtriangledown$ & $\bigtriangledown$ &  &  &
&  &  &  &  & \\
&  &  &  &  &  &  &  &  &  &  & $\bigtriangledown$ &  &  &  &  &  &  &  &
\\\hline
\end{tabular}
\caption{Arnold's tongue of resonances  ($\bigtriangledown$) in the $(\omega
,a)$-plane; $0.5\leq\omega\leq1.5$ (horizontally),  $0\leq a\leq
0.1$ (vertically).\label{key}}
%TCIMACRO{\TeXButton{E}{\end{table}}}%
%BeginExpansion
\end{table}%
%EndExpansion

We have demonstrated in \cite{ctm-port2} that depending on the parameters, the
system exhibits dynamical regimes that include a Hopf bifurcation followed by
classical Feigenbaum's cascade of period doubling leading to chaotic
pulsations, a Shilnikov-Hopf bifurcation etc.

For numerical simulations below we employed an explicit finite-difference
numerical code in the coordinate system attached to the interface
$\eta=x-s(t)$. Problem (\ref{he})-(\ref{jc}) is considered on a finite
interval $-L\leq\eta\leq L$ with the Dirichlet condition $u(-L,t)=0$
simulating the decay of the solution at $-\infty$ and Neumann condition
$\dfrac{\partial\theta}{\partial\eta}(L,t)=0$ reflecting stabilization of the
temperature far behind the interface in the burned matter (solid phase) .In
view of the fact that the dynamics of the problem requires a very fine
temporal resolution (to resolve sharp changes in the velocity), our experience
shows that there is no advantage in using implicit methods in this case (see
\cite{ctm-port2} for detail).

We note that in mathematical terms, the free boundary problem in
(\ref{he})-(\ref{jc}) governs the temporal evolution in the
infinitely-dimensional ``phase'' space of functions $u(x)$ and scalars $v$.
One way to present results would be through graphs of $u(x,t)$ versus $x,t$
and time histories $v(t)$ vs.~$t$. We also represent dynamics through
projections of the infinitely-dimensional phase space onto the 3-dimensional
space $[u(s(t)-1,t),u(s(t),t),u(s(t)+1,t)]$. I.e., the functional profile
$u(.,t)$ is represented by three values: at the interface and at two points
equidistant from it.

\section{Frequency locking}

The main objective of this paper is to demonstrate the phenomenon of frequency
locking for the driven free-boundary problem (\ref{he})-(\ref{jc}). To this
end we choose the initial mass concentration to be in the form of the
harmonically perturbed unity:
\begin{equation}
Z_{0}(s)=1+a\cos(\omega s) \label{concentr}%
\end{equation}
Note that unlike typical forced problems, the perturbation here is a function
of the dependent variable of the problem $s(t).$ Thus, strictly speaking, the
problem remains autonomous. Therefore the driving frequency $\omega$\ and the
frequency of the periodic solution, as seen in the power spectra and temporal
dynamics in the figures below, are not the same. Of course, it is possible to
introduce a new time $\tau=s(t)$, which would make the perturbation
''time''-dependent and the problem non-autonomous. However, it introduces a
stiff highly-variable coefficient $1/v$ at the time derivative of the heat
equation that makes numerical solution substantially more challenging.

Parameters in (\ref{arh1}) were selected to be $\sigma=0.1,$ $\alpha=4.5$
which correspond to a regime with simple periodic oscillations just past the
Hopf biurcation for the unperturbed problem. We vary the forcing amplitude $a$
from $a=0$ to $a=0.1$ with an increment of $0.005$ and the frequency $\omega$
from $\omega=0.5$ to $\omega=1.5$ with the step $\Delta\omega=0.05.$ The
results show \ typical structures of the Arnold's tongues in the $(a,\omega
)$-parameter plane. Note that for every pair of parameters the computation
involves numerical solving a nontrivial free-interface problem, and the
computational cost of an exhausting analysis of an extended domain in the
parameter plane becomes prohibitive.

Table 1 represents a crude outline of one of Arnold's tongues. We note that in
addition to the main tongue one can\ also observe what seems to be smaller
resonant tongues that correspond to higher order resonances. The resolution of
the computations however, does not allow us to make a definitive judgement
concerning additional tongues.

Next we discuss several representative solutions of the forced problem for
various parameter sets. Each of the Figures below contains a velocity profile
(top), its power spectrum\ (bottom-left), and a 3-dimensional projection of
the orbit into the space $[u(s(t)-1,t),u(s(t),t),u(s(t)+1,t)]$, where $u(.,t)$
is the temperature (bottom-right).

Fig. 1 depicts a simple periodic orbit. Fig.~2 shows a trajectory whose period
corresponds to twice the period of the forcing (note that the forcing has a
\textit{spatial} not temporal period). Figs.~3 and 4 depict periodic orbits of
high multiplicity in ultraharmonic and subharmonic domains respectively.
Fig.~5 demonstrates an example of what appears to be a chaotic trajectory,
which can be incurred from the power spectrum. This might indicate a presence
of a period-doubling cascade. Finally, Fig.~6 shows a typical quasi periodic
orbit. We remark that quasi periodic solutions with more than two basic
periods have also been observed.%
%TCIMACRO{\FRAME{dtbphFU}{6.045in}{3.5336in}{0pt}{\Qcb{Fig. 1. $a=.04$,
%$\omega=1$. Top: velocity profile; bottom-left: power spectrum; bottom-right:
%3-d projection of the orbit.}}{}{poli-res1a.eps}%
%width 6.045in;  height 3.5336in;  depth 0pt;  original-width 6.0174in;
%original-height 3.506in;  cropleft "0";  croptop "1";  cropright "1";
%cropbottom "0";  filename 'poli-res1a.eps';file-properties "XNPEU";}}}%
%BeginExpansion
\begin{center}
\includegraphics[
height=3.5336in,
width=6.045in
]%
{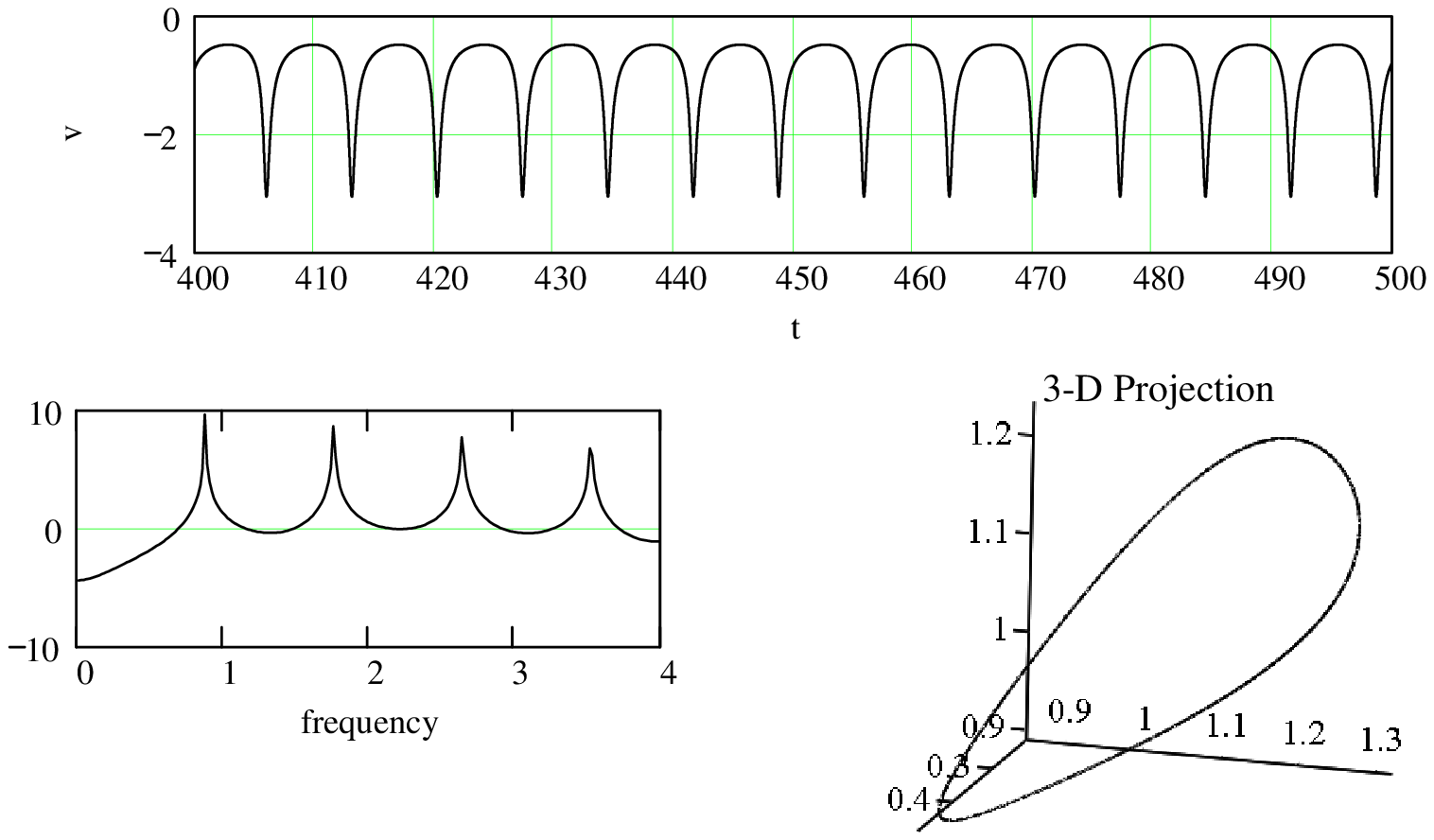}%
\\
Fig. 1. $a=.04$, $\omega=1$. Top: velocity profile; bottom-left: power
spectrum; bottom-right: 3-d projection of the orbit.
\end{center}
%EndExpansion%
%TCIMACRO{\FRAME{dtbphFU}{6.045in}{3.5336in}{0pt}{\Qcb{Fig.~2. $a=.08,$
%$\omega=1.4$}}{}{poli-res2a.eps}{\special{ language "Scientific Word";
%type "GRAPHIC";  maintain-aspect-ratio TRUE;  display "USEDEF";
%valid_file "F";  width 6.045in;  height 3.5336in;  depth 0pt;
%original-width 6.0174in;  original-height 3.506in;  cropleft "0";
%filename 'poli-res2a.eps';file-properties "XNPEU";}}}%
%BeginExpansion
\begin{center}
\includegraphics[
height=3.5336in,
width=6.045in
]%
{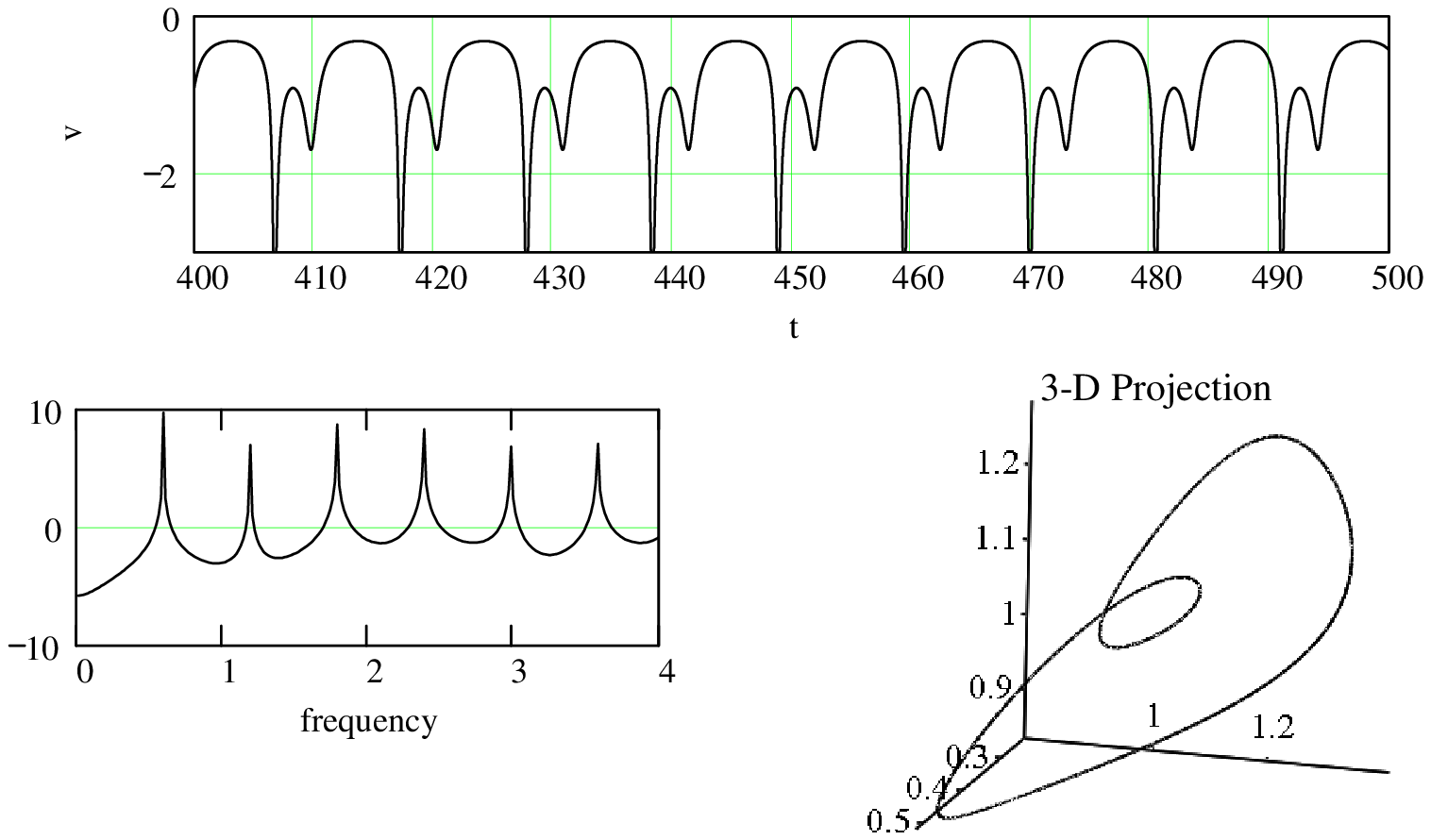}%
\\
Fig.~2. $a=.08,$ $\omega=1.4$%
\end{center}
%EndExpansion%
%TCIMACRO{\FRAME{dtbphFU}{6.045in}{3.5336in}{0pt}{\Qcb{Fig.~3. Ultraharmonic
%resonance of multiplicity 4 for $a=.08,$ $\omega=.65$}}{}{poli-res3a.eps}%
%width 6.045in;  height 3.5336in;  depth 0pt;  original-width 6.0174in;
%original-height 3.506in;  cropleft "0";  croptop "1";  cropright "1";
%cropbottom "0";  filename 'poli-res3a.eps';file-properties "XNPEU";}}}%
%BeginExpansion
\begin{center}
\includegraphics[
height=3.5336in,
width=6.045in
]%
{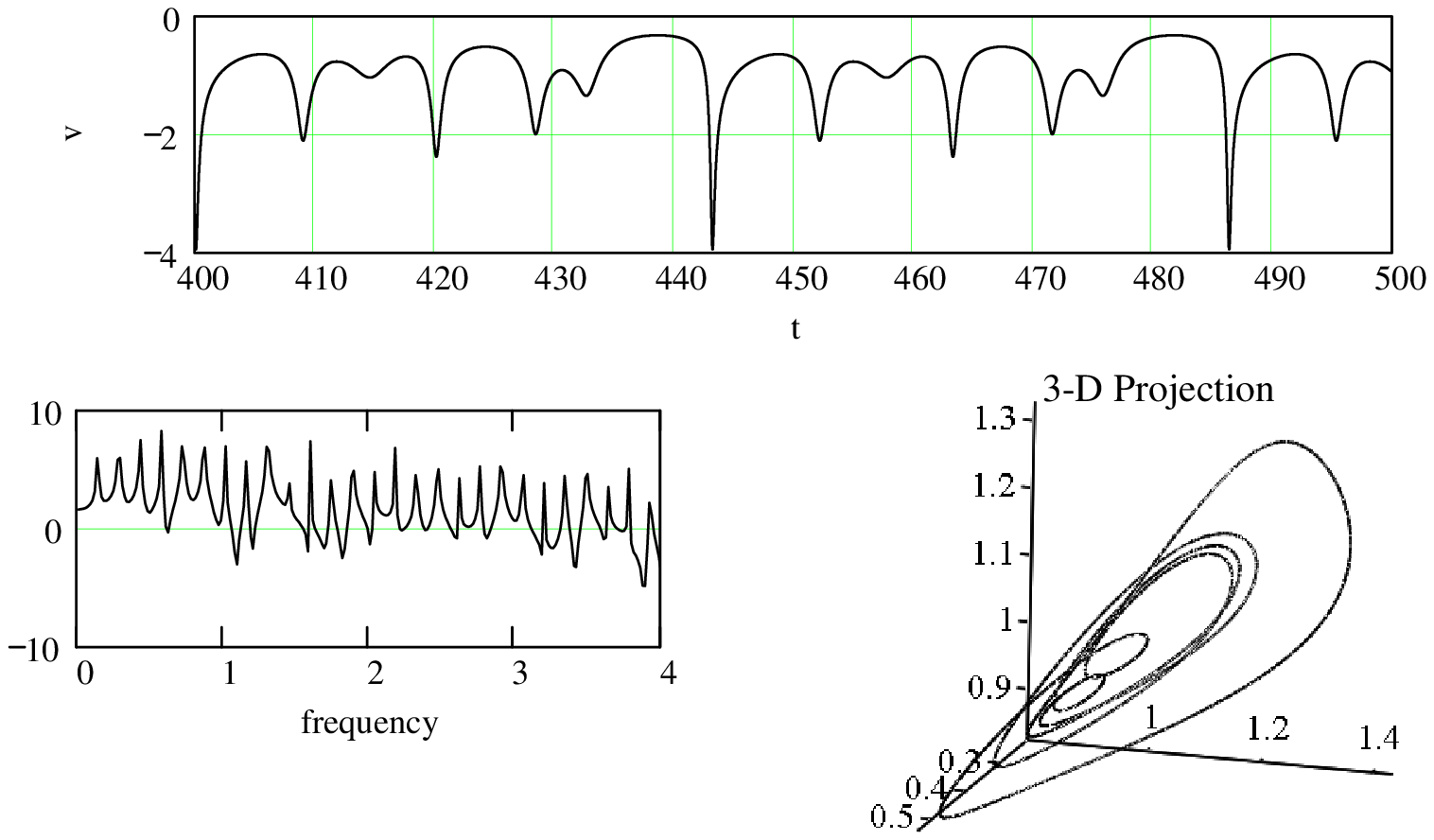}%
\\
Fig.~3. Ultraharmonic resonance of multiplicity 4 for $a=.08,$ $\omega=.65$%
\end{center}
%EndExpansion%

%TCIMACRO{\FRAME{dtbphFU}{6.045in}{3.5336in}{0pt}{\Qcb{Fig.~4. Subharmonic
%resonance of large multiplicity for $a=.08,$ $\omega=2.5$}}{}{poli-res4a.eps}%
%width 6.045in;  height 3.5336in;  depth 0pt;  original-width 6.0174in;
%original-height 3.506in;  cropleft "0";  croptop "1";  cropright "1";
%cropbottom "0";  filename 'poli-res4a.eps';file-properties "XNPEU";}}}%
%BeginExpansion
\begin{center}
\includegraphics[
height=3.5336in,
width=6.045in
]%
{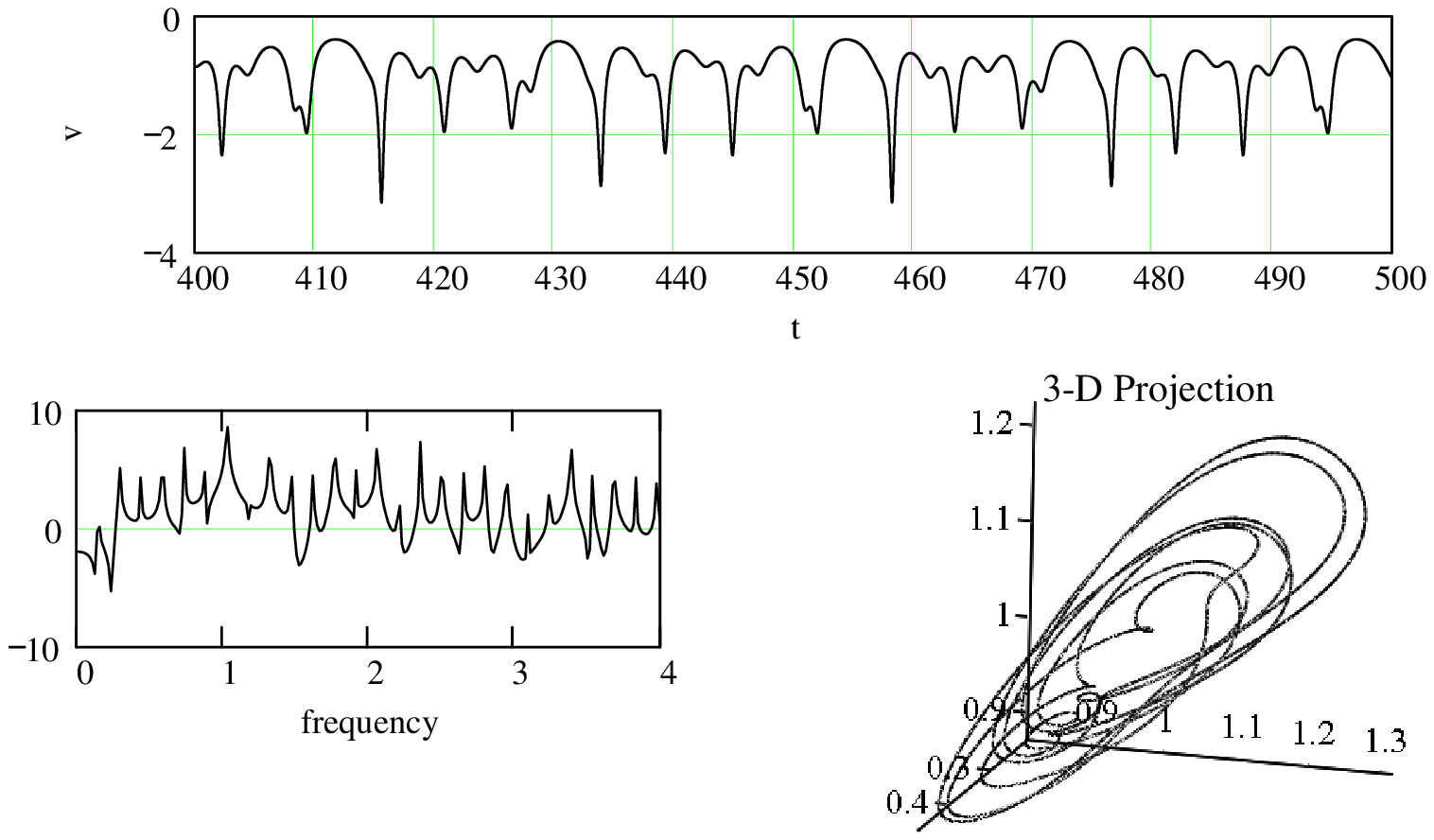}%
\\
Fig.~4. Subharmonic resonance of large multiplicity for $a=.08,$ $\omega=2.5$%
\end{center}
%EndExpansion%
%TCIMACRO{\FRAME{dtbphFU}{6.045in}{3.5336in}{0pt}{\Qcb{Fig.~5. Chaotic orbit
%for $a=.04,$ $\omega=1.4$}}{}{poli-res5a.eps}%
%width 6.045in;  height 3.5336in;  depth 0pt;  original-width 6.0174in;
%original-height 3.506in;  cropleft "0";  croptop "1";  cropright "1";
%cropbottom "0";  filename 'poli-res5a.eps';file-properties "XNPEU";}}}%
%BeginExpansion
\begin{center}
\includegraphics[
height=3.5336in,
width=6.045in
]%
{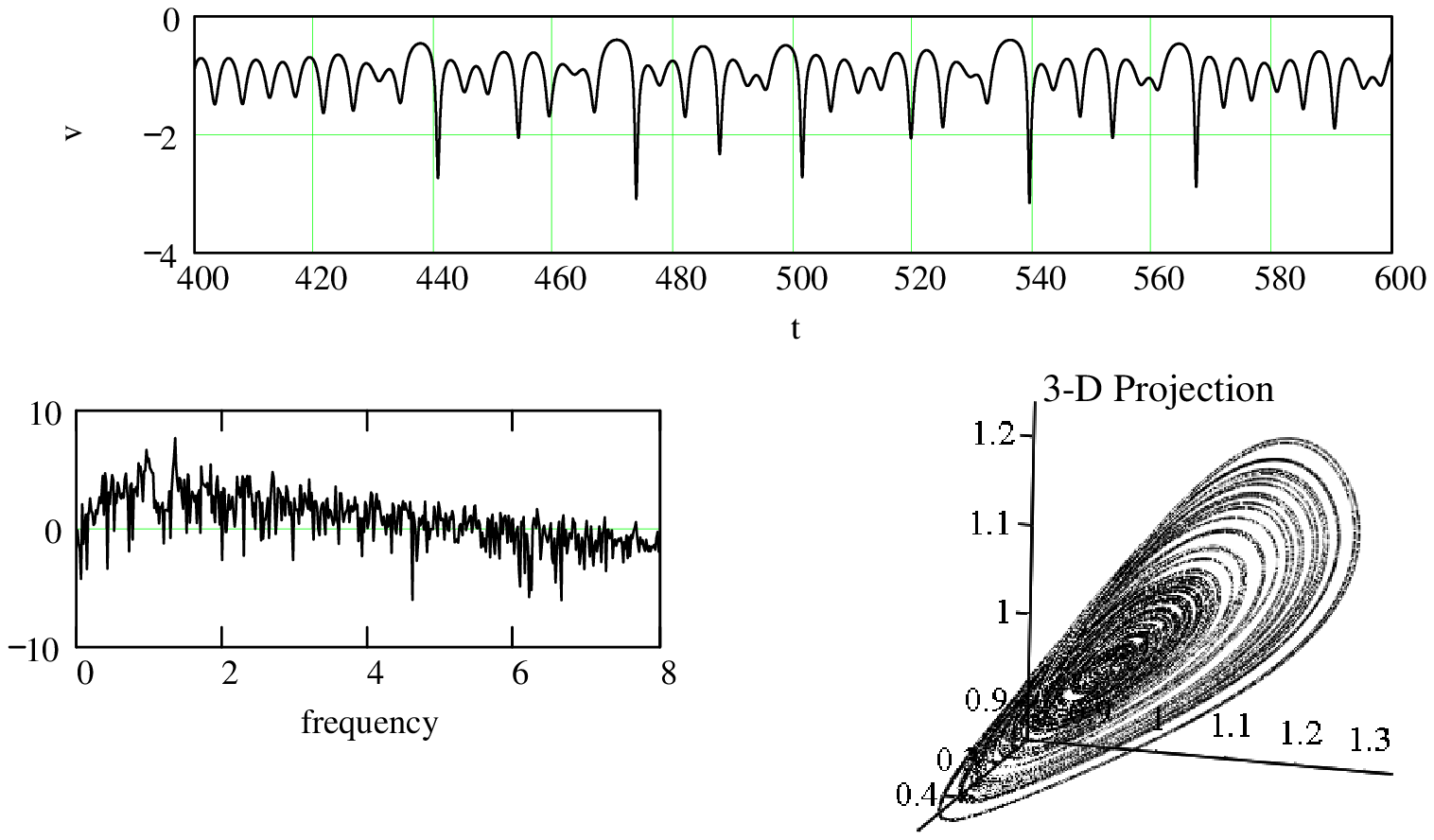}%
\\
Fig.~5. Chaotic orbit for $a=.04,$ $\omega=1.4$%
\end{center}
%EndExpansion%
%TCIMACRO{\FRAME{dtbphFU}{6.045in}{3.5336in}{0pt}{\Qcb{Fig.~6. Typical
%quasi-periodic orbit, $a=.1,$ $\omega=4$}}{}{poli-res6a.eps}%
%width 6.045in;  height 3.5336in;  depth 0pt;  original-width 6.0174in;
%original-height 3.506in;  cropleft "0";  croptop "1";  cropright "1";
%cropbottom "0";  filename 'poli-res6a.eps';file-properties "XNPEU";}}}%
%BeginExpansion
\begin{center}
\includegraphics[
height=3.5336in,
width=6.045in
]%
{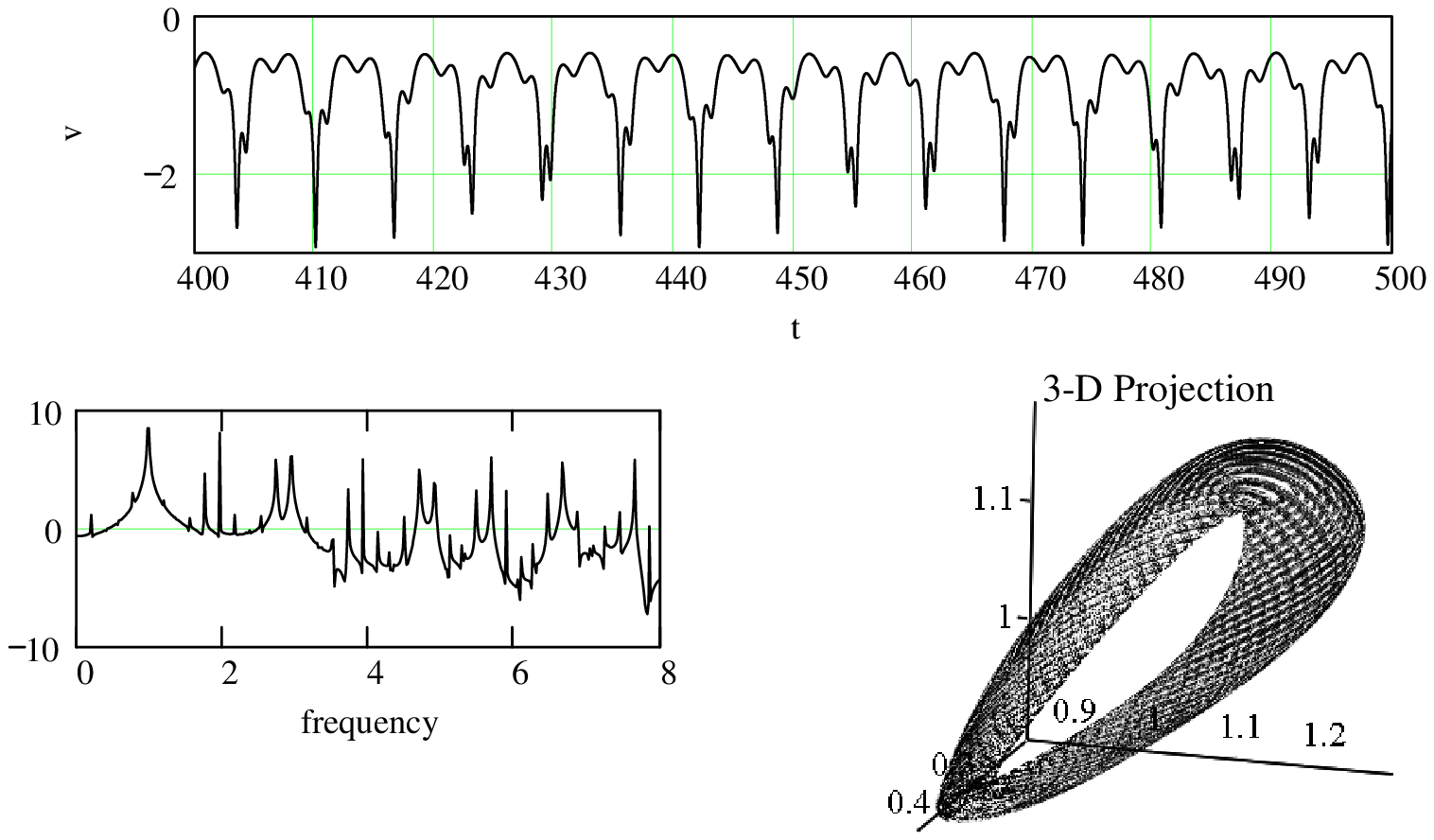}%
\\
Fig.~6. Typical quasi-periodic orbit, $a=.1,$ $\omega=4$%
\end{center}
%EndExpansion

\section{Concluding remarks}

Above we have demonstrated that the free-boundary problem (\ref{he}%
)-(\ref{jc}) exhibits some basic behavior quite analogous to the prototypical
finite-dimensional frequency locking systems such as, for instance, the driven
van der Pol equation. Since the sole objective of this letter is to
demonstrate the presence of the phenomenon of mode locking for the
free-interface problem we have deliberately chosen a very limited setting,
and, restrained ourselves from discussing a multitude of other questions that
arise in the context of this complex dynamical phenomenon.

There are a number of complex features observed in dynamical response of the
van der Pol system \cite{levi,mettin} and other finite-dimensional
counterparts that one would attempt to verify for our system . One would
naturally ask, for instance, whether the Arnold's tongues represent a dense
set with its complement forming a Cantor set in the $(a,\omega)$-plane,.
whether the Feigenbaum sequences occur within the tongues etc. It should be
noted that the dynamics of the unforced free-boundary problem varies
dramatically as the nonlinearity parameter $\alpha$ increases. In this letter
we considered only the simplest possible case corresponding to stable limit
cycle following Hopf bifurcation. It would be extremely interesting to
investigate how the frequency locking response changes with increasing
$\alpha.$ We hope to be able to pursue these issues in the near future.

\bigskip

\noindent\textbf{ACKNOWLEDGEMENT}. M.~F.~was supported in part by NSF Grant DMS-0207308.

\newpage
\end{document}